\documentclass{article}

\usepackage[preprint]{nips_2018}
\usepackage{amsmath}
\usepackage[ruled]{algorithm2e}
\usepackage{graphicx}

\usepackage[utf8]{inputenc} 
\usepackage[T1]{fontenc}    
\usepackage{hyperref}       
\usepackage{url}            
\usepackage{booktabs}       
\usepackage{amsfonts}       
\usepackage{nicefrac}       
\usepackage{microtype}      

\title{Revealing the Unobserved by Linking Collaborative Behavior and Side Knowledge}

\author{
  Evgeny Frolov\\
  Skolkovo Institute of Science and Technology\\
  \texttt{evgeny.frolov@skoltech.ru}
  \And
  Ivan Oseledets\\
  Skolkovo Institute of Science and Technology\\
  \texttt{i.oseledets@skoltech.ru}
}

\begin{document}

\maketitle

\begin{abstract}
    We propose a tensor-based model that fuses a more granular representation of user preferences with the ability to take additional side information into account. The model relies on the concept of ordinal nature of utility, which better corresponds to actual user perception. In addition to that, unlike the majority of hybrid recommenders, the model ties side information directly to collaborative data, which not only addresses the problem of extreme data sparsity, but also allows to naturally exploit patterns in the observed behavior for a more meaningful representation of user intents. We demonstrate the effectiveness of the proposed model on several standard benchmark datasets. The general formulation of the approach imposes no restrictions on the type of observed interactions and makes it potentially applicable for joint modelling of context information along with side data.
\end{abstract}

\section{Introduction}
User decision making process is influenced by various internal and external aspects, which in most cases are hardly observable and are difficult to collect. One of the greatest advantages of the collaborative filtering (CF) approach is that it does not require any specific knowledge about these aspects in order to generate recommendations. Particularly popular and successful representatives of the CF family, namely latent factor models,
help uncover general patterns from collective behavior, even if it is governed by a set of unidentifiable effects, events, motives, etc. Latent factor models describe these patterns in terms of a relatively small set of latent features learned from observations, which can be used to predict actual user preferences.

The CF approach relies on the assumption that collaborative information is sufficient, i.e. it accommodates all important variations in user behavior, so that intrinsic relations can be reliably learned from the data. This, however, may not always be the case. If the observed user-item interactions are too scarce, even latent factor models may fail to generalize well and tend to produce unreliable predictions \citep{zhang2014understanding, agarwal2009regression}. In the extreme case of the so called cold-start scenario \citep{ekstrand2011collaborative} such models simply become inapplicable without additional modifications.

One of the ways to deal with insufficient data and improve recommendation quality is to account for an additional knowledge about some of the observable aspects presumably associated with a hidden decision making mechanism. It can be, for example, user demographics, age, gender or item characteristics and properties. We will use the term \emph{side information} for this type of data.
The models that combine both collaborative data and side information are called \emph{hybrid} \citep{burke2002hybrid} and has been demonstrated to consistently improve performance of recommender systems in many cases, including high sparsity and cold-start scenarios.

A typical approach is to learn a latent factor model with additional constraints on the latent feature space induced by side information. An actual form of these constraints may take various forms, starting from regularization terms and simple linear transformations of latent factors to more intricate optimization objective expansions based on metric learning techniques and graph-based representations.
Such modifications push hybrid models towards more feasible and potentially more meaningful solutions. However, they also bring additional complexity, related to either optimization process itself or to hyper-parameter tuning.

One of the recent hybrid approaches called HybridSVD \citep{HybridSVD},
offers a very simple yet efficient model that goes a slightly different path.
It uses a generalized formulation of singular value decomposition (SVD) to enrich standard SVD-based approach called PureSVD \citep{Cremonesi2010} with side information. The model utilizes side information to measure how similar users or items are and virtually links them within collaborative data based on that similarity. This allows to uncover more valuable patterns that would otherwise stay unrecognized.
Notably, the model inherits the key benefits of its predecessor, such as a streamlined learning process with global convergence guarantees, deterministic output, simplified hyper-parameter tuning and an analytic form of \emph{folding-in}
computation \citep{ekstrand2011collaborative}, making it suitable even for highly dynamic online settings.

Nevertheless, as many other hybrid approaches, HybridSVD omits the question of an accurate user feedback representation. It can be a reasonable formulation when interaction data has the simplest form of an implicit feedback (e.g. likes or purchases). However, in a more general case user feedback has a more complex nature and often embodies several distinct types or modalities, which require careful treatment.
For example, an implicit feedback may split into different types of actions, such as click on a product page, placing an order or actual product purchase. Evidently, this corresponds to different levels of user engagement. Assigning appropriate weights to these actions in order to generate a single number (a.k.a. utility score) used in matrix-based formulations is a challenging empirical task. 
Even in the explicit case, such as rating values, the user feedback is better described in terms of ordinal relations rather than real numbers.
Indeed, from the fact that a user has assigned a 5-star rating to one movie and 2 stars to another, it does not follow that the user admires the former movie exactly 2.5 times higher than the latter. It only implies that the user prefers one movie to another and there are no arithmetic rules that allow to measure this difference.

The described challenges of proper feedback representation can be naturally addressed with the help of a tensor-based formulation
\citep{Coffee}, which gives us a versatile instrument to work with.
We propose a new hybrid tensor-based model that directly combines the key ideas of the aforementioned works by \cite{Coffee, HybridSVD} into a single general approach.
It allows to properly represent user preferences
and at the same time leverages side information in order to improve recommendations' quality and handle data sparsity. We provide efficient computational schemes for both offline learning and online recommendation generation in dynamic environments. We use several standard benchmark datasets to demonstrate our model's superiority to its predecessors.

\section{Problem formulation}
We start from a brief recap of both SVD-based and tensor-based models in order to introduce some common notation and prepare the ground for further generalization.

\subsection{Linking objects via side information}\label{subsec:hybridsvd}
The main idea of HybridSVD is to exploit the fact that standard SVD solves an eigendecomposition problem of scaled cosine similarity matrix. The corresponding scalar products between rows and columns in this view can than be replaced with more expressive and flexible bilinear forms. More formally, given a sparse matrix $A \in \mathbb{R}^{M \times N}$ that encodes interactions between $M$ users and $N$ items, the scalar product between its $i$-th and $j$-th rows can be replaced by:
\begin{displaymath}\label{eq:bilinear}
    c_{ij}  \sim \boldsymbol{a}_i^T S \, \boldsymbol{a}_j,
\end{displaymath}
where matrix $S \in \mathbb{R}^{N \times N}$ represents similarity between items (or their proximity) based on available side information and is responsible for virtually creating links between alike items even if they are never consumed together.
Likewise, the scalar products between columns of $A$ can be modified with the help of matrix $K \in \mathbb{R}^{M \times M}$, encoding side information-based similarity between users. Both $K$ and $S$ are required to be symmetric positive definite (SPD).

This leads to a generalized eigendecomposition problem, which has a solution in the form of standard truncated SVD of an auxiliary matrix:
\begin{equation}\label{eq:gsvd}
    \widehat{A} \equiv L_K^TAL_S^{} \approx \widehat{U}\Sigma\widehat{V}^T,
\end{equation}
where matrices $\widehat{U} \in \mathbb{R}^{M \times r}$ and $\widehat{V} \in \mathbb{R}^{N \times r}$ correspond to latent representation of users and items respectively in an auxiliary space; $r$ is a number of latent features and $\Sigma \in \mathbb{R}^{r \times r}$ is a diagonal matrix with $r$ elements above zero, sorted in descending order. Factors $L_S$ and $L_K$ are obtained from \emph{Cholesky decomposition} of similarity matrices, i.e. $S=L_S^{}L_S^T$ and $K=L_K^{}L_K^T$.

The model, however, is ``flat'' in a sense that \emph{it does not allow to distinguish between various types of feedback}.
As an example, if User A rates Item A with 2 out of 5 stars (negative preference) and User B rates Item B with 5 stars (positive preference), then high similarity between items A and B is unlikely to reflect shared tastes of the users. This, however, is quite opposite to what the model will actually learn and may create an undesired link.
This leads to inappropriate weighting of user feedback within the model and affects the resulting quality of recommendations.

\emph{We aim to resolve that issue in our model} and in order to do that we briefly describe the main idea of the Collaborative Full Feedback (CoFFee) model \citep{Coffee}, which allows to represent user feedback more appropriately, however, is not applicable for problems with side information.

\subsection{Higher order preference model}\label{subsec:tensorpref}
The CoFFee model encodes observed (\emph{user, item, feedback}) triplets into a sparse tensor of order 3, i.e. a multidimensional array with 3 distinct dimensions $\mathcal{A}{\,\in\,}\mathbb{R}^{M \times N \times F}$, where $F$ is a number of unique feedback values. For brevity, we will consider the case of only one type of feedback, such as a single 5-star likert scale with $F=5$ values along the 3-rd dimension. 
Generalization to higher order cases with several different scales or other types of feedback is trivial. Note, however, that $d$-dimensional problems with $d>4$ deserve a special care (see Section \ref{sec:related}).

The tensor is then approximated in the form of Tucker decomposition (TD) \citep{Kolda2009}, which can be viewed as a higher order generalization of SVD:
\begin{displaymath}\label{eq:tucker}
    \mathcal{A} \approx  \mathcal{G} \times_1U\times_2V\times_3W,
\end{displaymath}
where matrices $U \in \mathbb{R}^{M \times r_1}, V \in \mathbb{R}^{N \times r_2}$ have the same meaning as in the SVD case and newly introduced matrix $W \in \mathbb{R}^{F \times r_3}$ corresponds to the latent representation of user feedback. Symbol $\times_n$ stands for an $n$-mode product:
\begin{displaymath}\label{eq:n-mode-product}
    (\mathcal{A} \times_n X)_{i_1 \dots i_{n-1} \, j \, i_{n+1} \dots i_d} = \sum_{i_n} a_{i_1 \dots i_n \dots i_d} \, x_{ji_n}.
\end{displaymath}
Dense compressed tensor $\mathcal{G} \in \mathbb{R}^{r_1 \times r_2 \times r_3}$ is called the core of the decomposition and the tuple of numbers ($r_1, r_2, r_3$) is its \emph{multilinear rank}.

Similarly to SVD, the factors are required to have orthonormal columns. This allows to naturally extend the folding-in technique to higher order cases. Given a (sparse) matrix of only known user preferences $P \in \mathbb{R}^{N \times F}$, the matrix of predicted user preferences $\bar{P}$ with respect to all possible rating values can be estimated as:
\begin{equation}\label{eq:ho-folding-in}
    \bar{P} = VV^TPWW^T,
\end{equation}
which finalizes the necessary description part.
In the next section we introduce a new model that takes the best of both presented approaches in order to fuse side information with a higher order preference model.

\section{Proposed approach}\label{sec:proposed}
Following the same way SVD is generalized by Tucker decomposition, an auxiliary matrix $\widehat{A}$ from \eqref{eq:gsvd} can be generalized by an auxiliary tensor $\widehat{\mathcal{A}}$:
\begin{displaymath}\label{eq:cholesky_ten}
    \widehat{\mathcal{A}} \equiv \mathcal{A}\times_1L_K^T \times_2L_S^T \times_3L_R^T,
\end{displaymath}
where $L_R$ is a Cholesky factor of some SPD similarity matrix $R$ that corresponds to the third dimension.
With this formulation the model allows to naturally handle cases, similar to the example from Section \ref{subsec:hybridsvd}, by \emph{linking only items with the same feedback value}. This is achieved by setting $R=I$. The model, however, provides much more flexibility and allows to go beyond that scenario. In the presence of feedback similarity/correlation data (i.e. when $R$ is not just the identity matrix), \emph{the model allows to diffuse connections across feedback dimension} when it is required by the task or dictated by the structure of feedback data, e.g. when some feedback values are ``closer'' to each other in some sense. We will leave the discussion of its meaning for the later (see Section \ref{sec:future}).

The recommendation model is obtained from a low rank approximation of $\widehat{\mathcal{A}}$. As in the previous case, it can be achieved with the help of TD:
\begin{equation}\label{eq:hybrid-tucker}
    \widehat{\mathcal{A}} \approx  \mathcal{G} \times_1\widehat{U}\times_2\widehat{V}\times_3\widehat{W},
\end{equation}
where factor matrices are also required to have orthonormal columns. We call this model \mbox{\emph{HybridCoFFee}} to emphasize its ability to adequately represent higher order preference data and saturate it with side information.

Note that $\widehat{U} \in \mathbb{R}^{M \times r_1}, \widehat{V} \in \mathbb{R}^{N \times r_2}$ and $\widehat{W} \in \mathbb{R}^{F \times r_3}$ correspond to an auxiliary latent space. The latent representation of users, items and feedback in the original space is then given by
\begin{equation}\label{eq:auxfactors}
U = L_K^{-T}\widehat{U}, \quad V = L_S^{-T}\widehat{V}, \quad W = L_R^{-T}\widehat{W}.
\end{equation}
Columns of the resulting factor matrices satisfy $K$-, $S$- and $R$-orthogonality property, i.e. \mbox{$U^TKU = I_{r_1}$}, \mbox{\,$V^TSV = I_{r_2}$} and \mbox{$W^TRW = I_{r_3}$} ($I_r$ is an identity matrix of size $r$), which can be viewed as a constraint that \emph{structures the latent feature space according to real characteristics of modelled entities}.

The model also allows to control an overall contribution of side information into the learned latent representation as the similarity matrices are used in the form $K = I+\alpha K_0$, $S = I+\beta S_0$ and $R = I+\gamma R_0$, where zero-diagonal matrices $K_0, S_0$ and $R_0$ actually encode side information-based relations and $\alpha, \beta, \gamma$ are non-negative weighting parameters. Obviously, by setting $\alpha, \beta, \gamma$ to zero one gets standard CoFFee model.

Despite its similar look, the model has a few substantial differences from standard TD that require careful handling. In the next section we show how to efficiently compute it by a corresponding modification of the optimization objective. 

\subsection{Efficient computations}\label{subsec:compute}
A low rank approximation \eqref{eq:hybrid-tucker} can be obtained with a commonly used higher-order orthogonal iteration algorithm (HOOI), proposed by \cite{DeLathauwer2000best}. It solves the corresponding least squares problem by an alternating optimization procedure, where the objective is minimized with respect to one of the latent feature matrices while the other two are fixed. As shown by the authors of HOOI, the problem conveniently reduces to the following maximization task:
\begin{equation}\label{eq:hybrid-hooi}
    \max_{X}{\|\widehat{\mathcal{A}} \times_1\widehat{U}^T\times_2\widehat{V}^T\times_3\widehat{W}^T \|^2},
\end{equation}
where $X$ is picked iteratively from $\left\{\widehat{U}, \widehat{V}, \widehat{W}\right\}$ at each alternating optimization step; $\|\cdot\|$ denotes Frobenius norm, i.e. $\|\mathcal{A}\|^2=\sum_{i_1}\sum_{i_2}...\sum_{i_d}a^2_{i_1i_2...i_d}$.
Generally, the task can be efficiently solved by the means of SVD. Note, however, that unlike tensor $\mathcal{A}$, \emph{$\widehat{\mathcal{A}}$ is not necessarily sparse and may potentially blow up system resources}. In order to avoid its explicit formation we rewrite the inner term of \eqref{eq:hybrid-hooi} as
\begin{equation}\label{eq:hybrid-core}
    \widehat{\mathcal{A}} \times_1\widehat{U}^T\times_2\widehat{V}^T\times_3\widehat{W}^T \equiv
    \mathcal{A} \times_1U_K^T\times_2V_S^T\times_3W_R^T,
\end{equation}
where we utilize the multiplication properties of a series of matrices in the $n$-mode product \citep[Section 2.5]{Kolda2009} and use the substitution $U_K=L_K\widehat{U}, \, V_S=L_S\widehat{V}, \, W_R=L_R\widehat{W}$.

With the latter representation in \eqref{eq:hybrid-core} one can follow a standard technique to separate any factor matrix from the other two in order to perform an alternating optimization step. This is achieved by the virtue of \emph{tensor unfolding} \citep[Section 2.4]{Kolda2009}.
For example, to optimize for $\widehat{U}$ we arrive at the following expression:
\begin{displaymath}
    \max_{\widehat{U}}{\|\widehat{U}^T L_K^T A^{(1)} \left(W_R \otimes V_S\right)\|^2},
\end{displaymath}
where matrix $A^{(i)}$ \emph{denotes a mode-$i$ unfolding of} $\mathcal{A}$ and $\otimes$ stands for Kronecker product. The corresponding solution is then given by the leading left singular vectors of $L_K^T A^{(1)} \left(W_R \otimes V_S \right)$. Similar transformations along modes 2 and 3 give the update rules for the rest of the factors. See Algorithm \ref{alg:hybrid-hooi} for a full description of the optimization process.

Note that the product $A^{(1)} \left(W_R \otimes V_S \right)$ has the same structure as in the standard TD case. Therefore, for moderately sized problems it can be computed without explicitly forming $W_R \otimes V_S$ by performing a series of matrix multiplications with unfolded tensors \citep{andersson1998improving}. For larger problems the memory bottleneck induced by intermediate computation results can be circumvented by iteratively updating entries of the final result in a simple nested loop instead of performing matrix multiplications.
    \begin{algorithm}[t]\label{alg:hybrid-hooi}
      \SetKwInOut{Input}{Input}
      \SetKwInOut{Output}{Output}
      \Input{\quad Tensor $\mathcal{A}$ in sparse COO format,\\
      \quad Tensor decomposition ranks $r_1, r_2, r_3$, \\
      \quad Cholesky factors $L_K, L_S, L_R$}
      \Output{\quad $\mathcal{G}, \widehat{U}, \widehat{V}, \widehat{W}$}
      Initialize $\widehat{V}, \widehat{W}$ by random matrices with orthonormal columns.\\
      Compute $V_S=L_S\widehat{V}, W_R=L_R\widehat{W}$.\\
      \Repeat{norm of the core ceases to grow or exceeds maximum iterations}{
      $\widehat{U} \leftarrow$ $r_1$ leading left singular vectors of $L_K^T\, A^{(1)} \left(W_R \otimes V_S \right)$\\
      $U_K \leftarrow L_K\widehat{U}$\\
      $\widehat{V} \leftarrow$ $r_2$ leading left singular vectors of $L_S^T\, A^{(2)} \left(W_R \otimes U_K \right)$\\
      $V_S \leftarrow L_S\widehat{V}$\\
      $\widehat{W}, \, \Sigma, \, Z \leftarrow$ $r_3$ leading singular triplets of $L_R^T\, A^{(3)} \left(V_S \otimes U_K \right)$\\
      $W_R \leftarrow L_R\widehat{W}$\\
      $\mathcal{G} \leftarrow$ reshape matrix $\Sigma Z^T$ into shape ($r_3, r_1, r_2$) and transpose
      }
      \caption{Practical algorithm for hybrid HOOI}
    \end{algorithm}
\paragraph{Online recommendations.}
As in the case with CoFFee or HybridSVD the orthogonality of columns in factor matrices allows to derive an efficient expression for higher-order hybrid folding-in.
In the user case, it helps to solve the problem of recommendations for unrecognized or newly introduced users with only a few known preferences. Likewise, in the item case it allows to quickly find item representation in the latent space based on a few interactions with it.
As an example, the following expression is a generalization of the tensor folding-in to the hybrid case, which allows to estimate new user preferences (c.f. \eqref{eq:ho-folding-in}):
\begin{equation}\label{eq:ho-hybrid-folding-in}
    \bar{P} = V^{}V_S^TPW_R^{}W^T,
\end{equation}
where $V$ and $W$ are defined according to \eqref{eq:auxfactors}.
This allows to avoid recomputing the whole model in response to frequent system updates. It is especially viable in highly dynamic online environments, where users expect an instant response from recommendation services and/or new items arrive rapidly. In our experiments we use this formula to generate recommendations for known users as well.
\paragraph{Rank truncation.}
Hyper parameter-tuning can be a tedious task. Unlike many other approaches, SVD-based methods provide a luxury of minimal hyper parameter tuning via simple rank truncation of latent factors. Having computed the model of rank $r$ one can easily find a reduced model of any rank $r'<r$ by truncating its factors to the first $r'$ components.
Even though it is not directly applicable in the tensor case, it is still possible to avoid redundant computation of the model with lower multilinear rank values by the means of \emph{tensor rounding} technique.
More formally, given some factor matrix $X \in \{\widehat{U}, \widehat{V}, \widehat{W}\}$, which corresponds to some mode $i \in \{1, 2, 3\}$, and a new rank value $r < \text{rank}(X)$, the first step is to compute $r$ leading singular triplets $U_r, \Sigma_r, V_r$ of the unfolded core $G^{(i)}$. Then the new factor matrix $X_r$ of the reduced rank $r$ is calculated as $X_r = X U_r$ and the new truncated core $\mathcal{G}_r$ is obtained by reshaping matrix $\Sigma_r^{} V_r^T$ back to the tensor of order 3 with conforming size. Note that due to typically small multilinear rank values finding SVD of an unfolded core is computationally cheap.


\section{Evaluation methodology}
We conduct 5-fold cross-validation (CV) experiment for standard top-$n$ recommendation scenario performing splits by users. At every fold we randomly mark 20\% of users that were not yet tested. We \emph{randomly} hide 10 consumed items of every marked user to form the \emph{holdout set}. This allows to have both high and low ratings in the holdout and, therefore, to \emph{evaluate recommendations against both negative and positive user preferences}.
User feedback is considered to be positive if the rating value is equal or above 4 with the top rating being 5.
The remaining items from the marked users as well as all the preferences of 80\% of unmarked users form the \emph{training set}. At each fold we generate recommendations for the marked users and evaluate them against the holdout. CV results are averaged and reported along with 95\% confidence intervals based on the paired $t$-test criterion.

\paragraph{Metrics.}
As has been shown by \cite{Coffee}, standard evaluation metrics exhibit a \emph{positivity bias}, i.e. only consider the performance in terms of how relevant recommended items are and completely disregard how likely it is to get recommended something irrelevant. The latter, however, may have a dramatic impact on the perceived quality of a recommendation service and affects user retention.
In order to account for such effects we report not only the scores for standard relevance- and ranking-based metrics, but also evaluate models against the \emph{normalized Discounted Cumulative Loss} (nDCL), proposed by the same authors. It serves as a proxy measure for user disappointment and estimates how likely is a user to remain unsatisfied with provided recommendations. As it follows from the name, nDCL is the opposite of standard normalized Discounted Cumulative Gain (nDCG) metric. Note that \emph{models with similar nDCG may have different nDCL score}.

\paragraph{Datasets}
We use 3 standard benchmark datasets: \emph{MovieLens-1M} (ML1M), \emph{MovieLens-10M} (ML10M), and \emph{BookCrossing} (BX), published by Grouplens\footnote{https://grouplens.org/datasets/}. These datasets have very different levels of data sparsity and therefore allow to examine how sensitive our model is to the lack of collaborative information in comparison to other models. We do not perform any special preprocessing for the Movielens datasets. In the BX case we filter out users with more than 1000 ratings as they are unlikely to represent real consumption patterns. We also remove books with only one rating provided by a single user as unreliable. Ratings in the BX dataset range from 1 to 10. In order to have uniform representation across all datasets, we divide them by 2, giving a range from 0.5 to 5 with 0.5 step, similarly to ML10M. Ratings in the ML1M dataset are integer values from 1 to 5.

\paragraph{Algorithms.}
We compare our method to both CoFFee and HybridSVD approaches. We additionally use standard baseline models, namely PureSVD \citep{Cremonesi2010}; a heuristic model that recommends items based on their aggregated similarity to known user preferences (CB); and a non-personalized model that simply recommends the most popular items (MP).
Models are tuned on the first CV fold and the best found configuration corresponding to the highest nDCG score is then used across the remaining folds. In the case of PureSVD the only varying hyper-parameter is the rank of SVD. In the CoFFee model we tune its multilinear rank with the requirement for mode-1 and mode-2 ranks to be always equal and take values from the same range as the rank of PureSVD. Mode-3 rank takes values from \{2, 3, 4\}.
In the HybridSVD case we firstly tune its rank with a fixed weight value for side information set to 0.5. After an optimal rank is found we perform additional evaluation to find the most suitable weight value from \{0.1, 0.5, 0.9\}. Similar procedure is performed for HybridCoFFee with the same requirement on rank values as for the CoFFee model. SVD-based models use rank truncation to avoid redundant calculations during rank tuning. Likewise, tensor-based models use tensor rounding.

\paragraph{Side information.}
We used the information from TMDB database\footnote{https://www.themoviedb.org} to complete movie data in the Movielens datasets with information about \emph{cast, directors} and \emph{writers} along with already present \emph{genre} information. BX dataset provides additional information about \emph{authors} and \emph{publishers}.
There is no additional information about users or ratings, which renders $L_K$ and $L_R$ to be simply identity matrices.
For each dataset we \emph{inclusively merge all side data} by independently constructing similarity matrices $S_i$ for each particular feature $i$ and then combining them into a single similarity matrix with a simple summation $\frac{1}{n_f}\sum_{i=1}^{n_f} S_i$, where $n_f = 4$ in the Movielens case and $n_f = 2$ in the BX case. We used the same similarity measures
for constructing $S_i$ as in the HybridSVD paper.

\begin{figure}[t]
\centering{
\includegraphics[width=138mm]{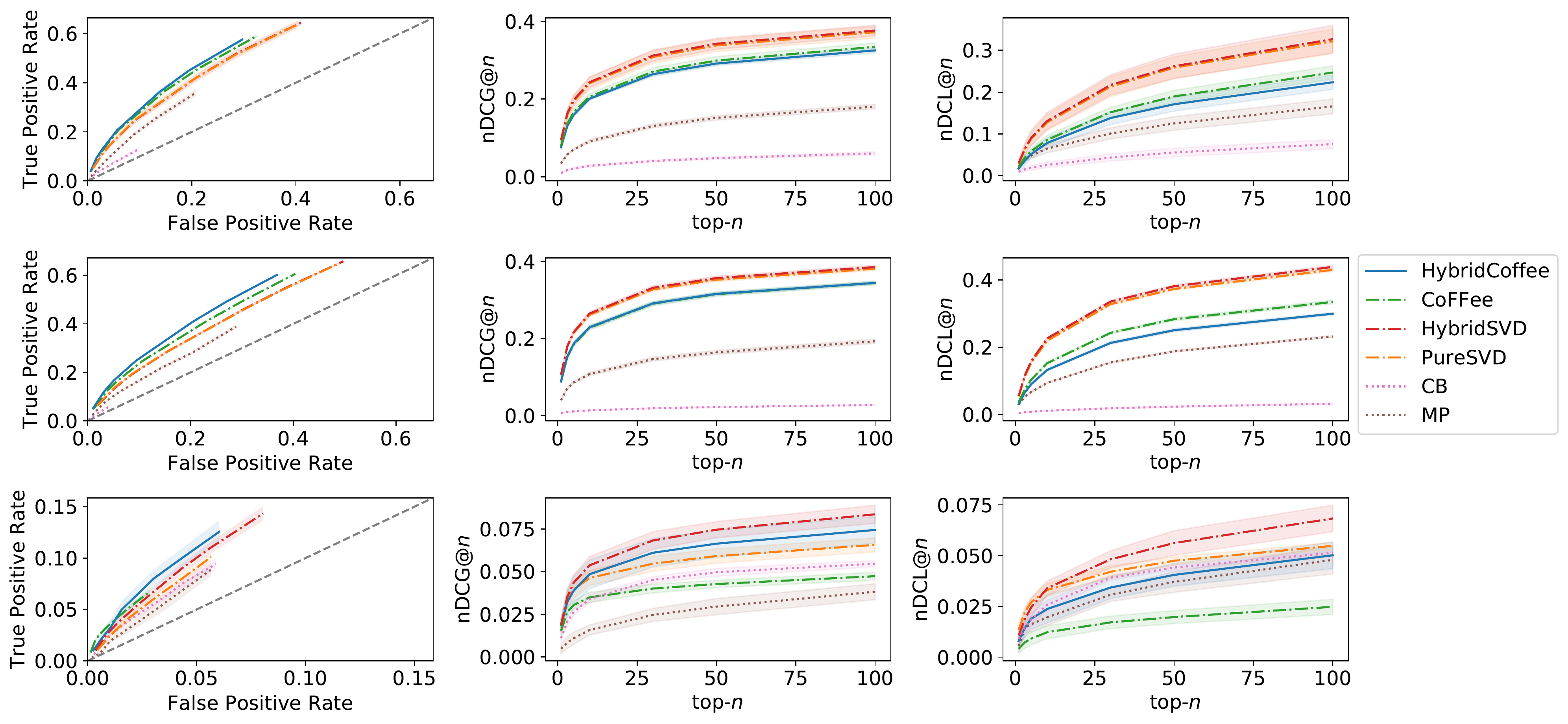}
}
\caption{The ROC curves (1st column), nDCG@$n$ (2nd column) and nDCL@$n$ (3rd column). Rows correspond to different datasets: 1st row for ML1M, 2nd row for ML10M, 3rd row for BX.
For the first 2 columns the higher the curve, the better, for the last column the lower the curve, the better. Shaded areas show a confidence interval.}
\label{fig:results}
\end{figure}

\section{Results}
We report 3 key evaluation metrics for all three datasets, that allow to assess the quality of recommendation models: an overall ratio of relevant recommendations to irrelevant, measured by Reliever Operator Characteristic curve (ROC), position of \emph{relevant} predictions in top-$n$ recommendation list, measured by nDCG and position of \emph{irrelevant} predictions in top-$n$ recommendation list measured by nDCL (see Figure \ref{fig:results}). Note, that there is typically some balance between high relevance of recommendations and high probability to generate irrelevant recommendations as well.

In order to correctly interpret results it is important to note, that low nDCG scores do not necessarily mean low quality of recommendations. If a model with low nDCG produces high enough ROC curve and at the same time shows low nDCL it simply means that the model makes more ``safe'' recommendations. Instead of recommending something irrelevant it pushes to the top more of unrated items, which is generally a better strategy. In contrast, if the relevance-based scores as well as nDCL score are all low, it indicates a poor performance.

For example, as can be seen from the first row of Figure \ref{fig:results}, both CB and MP models have low nDCL, however, their relevance-based scores are also low, which means that these models provide unsatisfactory recommendations.
In contrast, HybridSVD provides the highest (or one of the highest) nDCG score in general. However, it also pushes one of the highest numbers of irrelevant items to the top of recommendations list, as indicated by its nDCL score. As has been argued in Section \ref{subsec:hybridsvd}, this is likely to be the result of unreliable connections, created by the model, between items with very different rating values.

As it follows from the results, HybridCoFFee \emph{outperforms all other models in terms of the proportion of relevant recommendations to irrelevant ones}. Its advantage is especially vivid in the second row of the figure, which corresponds to the ML10M dataset. On this data our model is able to decrease nDCL score below the standard CoFFee model, while keeping nDCG score at the same fairly high level. This decrease in of irrelevant recommendations is immediately reflected by the ROC curve.

Generally, our model exhibits the best balance between the key 3 evaluation aspects. It does not suffer from the sparsity of data as, for example, the tensor-based CoFFee model in the BX case (see the ROC curve on the third row of the Figure \ref{fig:results}). It maintains high relevance of recommendations and \emph{generates more safe predictions}, allowing to avoid potential user disappointment.


\section{Related}\label{sec:related}
As we have demonstrated, our work is based on a generalization of two models, namely HybridSVD and CoFFee.
There many other factorization techniques that allow to achieve similar functionality. One of the most well-known tensor-based models, called Multiverse \citep{Karatzoglou2010}, also uses TD format, however, for a different data representation. The authors of Multiverse propose to encode any data, including side information, within additional dimensions. They also propose to seek for a solution to the corresponding optimization problem with the help stochastic gradient descent and provide efficient scheme for computations.

This model, however, can be hardly applied to the problems with many dimensions, as the storage required for TD factors depends exponentially on the number of dimensions. This leads to the so called \emph{curse of dimensionality} problem. More appropriate tensor formats for multidimensional cases of a higher order would be Tensor Train (TT), proposed by \cite{Oseledets2011}, or Hierarchical Tucker (HT), proposed by \cite{grasedyck2010hierarchical}.

The curse of dimensionality can also be avoided with Candecomp/Parafac decomposition (CP). The TAPER model proposed by \cite{ge2016taper} uses CP as a workhorse for a unified representation, where various sources of information are glued together with the help of additional regularization constraints. It also imposes additional locality constraints, requiring similar entities to be close to each other in the latent space. Note, however, that CP decomposition is generally unstable and may require additional efforts in order to ensure convergence. Moreover, in a general case it does not impose orthogonality constraint on the columns of factor matrices. This leads to a more complicated folding-in procedure that requires additional optimization steps.

There are many more regularization-based models that allow to impose a desired structure on the latent feature space. Such models are based on a class of methods often called \emph{collective} \citep{singh2008relational} or \emph{coupled} factorization \citep{rafailidis2016modeling, barjasteh2015cold}.
We would like to emphasize, however, that unlike the majority of hybrid factorization methods, our approach provides a simple computational framework that \emph{uses SVD as an atomic operation} and also provides the simplest possible form of the folding-in technique, which does not require any extra optimization steps. Wide availability of robust and highly optimized implementations of SVD in different programming languages adds additional practicality points to the proposed approach.

\section{Discussion and future work}\label{sec:future}
We have presented a tensor-based approach that combines the ability to more adequately model user preferences and allows to incorporate side knowledge in order to handle data sparsity and improve the quality of recommendations. Based on the evaluation results we show that the proposed model provides the best balance between providing good recommendations and avoiding undesired user disappointment.

Note, that the general formulation of our approach allows to handle context information, such as time, place, mood, situation, etc., in within additional dimensions, similarly to various types of feedback. This can be an interesting direction for further research, especially in the cases where context contains additional information about correlations between its different values. The key benefit of the model in that case is that it would allow to handle even more extreme sparsity levels induced by multidimensional representation.

Based on the remark about the curse of dimensionality problem of TD, another interesting direction for research is applying the key ideas presented in this work to more appropriate tensor formats such as TT or HT.

\bibliographystyle{ACM-Reference-Format}
\bibliography{bibliography}

\end{document}